# Bullet Retarding Forces in Ballistic Gelatin by Analysis of High Speed Video

Steven Gaylord, Robert Blair, Michael Courtney
U. S. Air Force Academy,[1] 2354 Fairchild Drive, USAF Academy, CO, 80840
Michael_Courtney@alum.mit.edu

Amy Courtney
BTG Research, PO Box 62541 Colorado Springs, CO 80920
amy_courtney@post.harvard.edu

**Abstract**
Though three distinct wounding mechanisms (permanent cavity, temporary cavity, and ballistic pressure wave) are described in the wound ballistics literature, they all have their physical origin in the retarding force between bullet and tissue as the bullet penetrates. If the bullet path is the same, larger retarding forces produce larger wounding effects and a greater probability of rapid incapacitation. By Newton's third law, the force of the bullet on the tissue is equal in magnitude and opposite in direction to the force of the tissue on the bullet. For bullets penetrating with constant mass, the retarding force on the bullet can be determined by frame by frame analysis of high speed video of the bullet penetrating a suitable tissue simulant such as calibrated 10% ballistic gelatin. Here the technique is demonstrated with 9mm NATO bullets, 32 cm long blocks of gelatin, and a high speed video camera operating at 20,000 frames per second. It is found that different 9mm NATO bullets have a wide variety of potential for wounding and rapid incapacitation. This technique also determines the energy transfer in the first 15 cm and/or first 30 cm of tissue, which are important parameters in estimating the probability of rapid incapacitation in some of the ARL/BRL models. This method predicts that some 9mm bullets have a much higher probability of rapid incapacitation than others and the rank ordering of bullet effectiveness is in agreement with other studies.

## Introduction

Quantifying the potential for wounding and rapid incapacitation of a given bullet in tissue has always been of considerable importance from both offensive and defensive standpoints. For many years the U.S. Army (BRL, now ARL) studied terminal ballistics in 20% ballistic gelatin. In the last 20 years, most agencies have transitioned to calibrated 10% ballistic gelatin at a specified temperature because it has been shown that both penetration depth and wound cavities closely match observations in living tissue. The most common means to quantify observations in ballistic gelatin are the penetration depth, the permanent cavity volume, and the temporary cavity volume (Krauss, 1960; Courtney and Courtney, 2012). As the importance of the ballistic pressure wave has been illuminated (Courtney and Courtney, 2007a), it is clear that one can also insert a high speed pressure transducer into gelatin to record pressure transients. One can also estimate the ballistic pressure wave from the retarding force.

Carroll Peters (1990) pointed out that all three of the recognized wounding mechanisms have their origin in the retarding force between bullet and tissue. The permanent cavity is created by an intense stress field in the immediate vicinity of the penetrating projectile. These stress waves decay rapidly with distance from their origin so the region of "prompt damage" tends to be close to the projectile path through tissue. The permanent cavity may be enlarged if the tissue is stretched beyond the elastic limit by the temporary cavity. The temporary cavity arises because the retarding force accelerates tissue which then stretches until the combination of inertia, weight, and elasticity causes it to spring back into place. Inelastic tissues such as liver, spleen, and brain stretch much less than elastic tissues such

---

[1] Distribution A. Approved for public release. Distribution unlimited. The views expressed in this paper are those of the authors and do not necessarily represent those of the U.S. Air Force Academy, the U.S. Air Force, the Department of Defense, or the U.S. Government.

# Bullet Retarding Forces in Ballistic Gelatin by Analysis of High Speed Video

muscle and lung and suffer more damage from being stretched. The retarding force also gives rise to a ballistic pressure wave which can propagate and cause wounding at locations distant from the bullet path. (Courtney and Courtney, 2007; Suneson et al. 1990a, 1990b; Krasja, 2009; Selman et al., 2011)

High speed video is now in common use to study bullet impacts in ballistic gelatin, though it does not seem common to go beyond visualization and perhaps determining the dimensions of the permanent cavity and temporary cavity. Quantifying the retarding force and its importance in wound ballistics probably originated with the Swedes in the work of Bo Janzon and his colleagues (Janzon, 1983; Bellamy and Zajtchuk, 1990). There have been hints that similar techniques may have been used (perhaps are still currently used) by the Ballistics Research Laboratory (later ARL) and possibly also by the FBI ballistics research facility, but (to our knowledge) detailed techniques and or results have not been published. Early work was done with flash x-ray to determine the bullet position as a function of time. Flash x-ray has the advantage that it can be used in media such as living tissue, Swedish soap, and other opaque materials. High speed video is now more cost effective now and more widely available and can be used with translucent materials such as ballistic gelatin. (See Bruchey and Sturdivan, 1968.)

The analysis principles are the same as in most use of video for kinematic analysis. The position of the object of interest (bullet) is determined frame by frame in the coordinate system of pixels which is then converted to the length unit by use of an appropriate scale. The change in position with respect to time is the velocity. The change in velocity with respect to time is the acceleration. If suitable conditions are met, then other parameters of interest such as kinetic energy and forces can be determined by the elementary laws of physics. This paper describes the details that can be employed for kinematic analysis of bullets penetrating ballistic gelatin when captured on high speed video with a suitable frame rate and pixel resolution. This technique is then demonstrated by determining retarding force vs. penetration depth for four different 9mm NATO bullets. Conventional wound profiles (permanent cavity and temporary cavity) are also determined. If other parameters such as $E_{15}$ and $E_{30}$ (the energy deposit in the first 15 cm and the first 30 cm, respectively) are also desired (Neades and Prather, 1991) then they can be gleaned from the analysis.

**Method**
Ballistic gelatin was prepared to a 10% concentration and calibrated per the FBI protocol. Each trial was recorded with an IDT Motion Pro X4 high-speed camera at 20,000 frames per second. The camera position and lens were adjusted for a field of view approximately 15 cm high and 60 cm wide centered on the ballistic gelatin. A transparency sheet with printed scale with tick marks every 2.5 cm over a distance of 25 cm was placed on the gelatin. This scale was used to calibrate the horizontal distance so that the horizontal position of the bullet could be determined. The video was then analyzed frame by frame and the position of the bullet is recorded for each frame.

A spreadsheet was created with columns for time (shifted for impact at t = 0 s), horizontal position (in pixels), and horizontal position (in feet). A measured velocity column (ft/s) was created where the velocity was computed as the change in position from the last frame to the current frame divided by the change in time. At 20,000 frames per second, the change in time was constant: 0.00005 s.

The directly measured velocity column has several experimental sources of noise and does not decrease monotonically as the physical considerations demand. The experimental noise is dominated by discretization error. The measurement technique will suggest that the bullet has moved an integer number of pixels between frames, 5 to 15 pixels is common. Even when this is converted to feet, the change in position only takes on one of a small number of fixed, discrete values between frames. Thus the velocity measured between any two frames only takes on one of a small number of fixed, discrete values. Of course, the real velocity is decaying continuously. Other sources of experimental noise are imperfect optics, and changes in how the gelatin refracts light as it deforms with the passing bullet.

# Bullet Retarding Forces in Ballistic Gelatin by Analysis of High Speed Video

To smooth out the noise, the measured velocity vs. time data is fit to a rational model. Polynomial models can be used and work well in some cases, but rational models are generally better behaved and easier to enforce having a known initial value (the impact velocity determined with an optical chronograph) and a monotonic decay that asymptotically approaches zero. Here, a rational function is employed with the form:

$$V(t) = \frac{V_0}{1 + a_1 t + a_2 t^2 + a_3 t^3},$$

where V(t) is the model velocity as a function of time, $V_0$ is the impact velocity determined from an optical chronograph, and $a_1$, $a_2$, and $a_3$ are parameters determined from a non-linear least squares fit of the model function to the measured velocity data. In principle, $V_0$ could also be treated as an adjustable parameter, but an optical chronograph determines the impact velocity much more accurately (0.3%) than allowed by the resolution and frame rate of the video camera, and non-linear regression to rational functions is better behaved with fewer parameters. The authors also favor empirical models with the proper limiting behavior.

MS Excel and other spreadsheet programs (the authors favor Libre Office Calc) often have a few built in functions like exponentials and polynomials but do not have built in facilities for non-linear regression to rational functions. A polynomial function can often be made to work with a few caveats: 1) Polynomials work better if you force the vertical intercept to be equal to the initial velocity. (Set the constant term equal to the initial velocity.) 2) There is often some trial and error involved in selecting a proper number of terms so the polynomial decreases monotonically. A fourth degree polynomial is a reasonable first choice, but sometimes the degree needs to be lowered to suppress local maxima. 3) You may need to put in V = 0 by hand in the measured data for a few frames to encourage the model to approach the vertical axis more like an asymptote (the acceleration is also going to zero as the bullet stops).

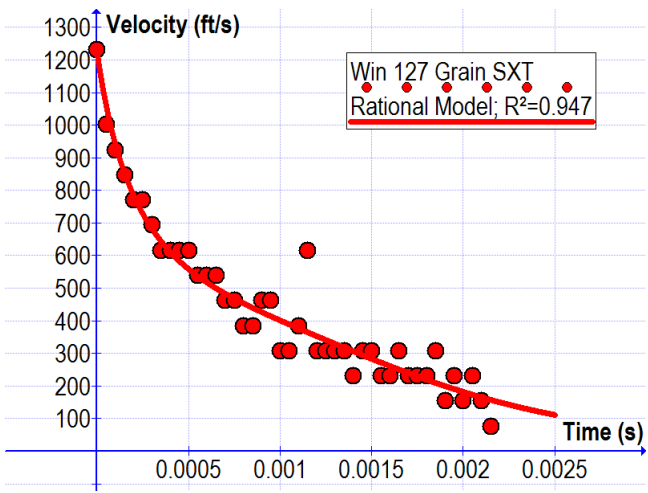

Figure 1: Measured and model velocities in ballistic gelatin for the 127 grain Winchester Ranger SXT in 9mm NATO impacting at 1232 ft/s.

Rather than work around the caveats in using a polynomial, the authors prefer to use a rational function. The measured velocity vs. time data is transferred into a program with non-linear regression capabilities (SciDaVis, Gnuplot, and Graph.exe are three favorites), and a non-linear least-squares fit is performed with the rational model. Once the model parameters are determined, the formula is copied back into the

# Bullet Retarding Forces in Ballistic Gelatin by Analysis of High Speed Video

spreadsheet to create a column for the model velocity (ft/s) at the time corresponding with each frame. Figure 1 shows the measured and model velocities for the 127 grain Winchester Ranger SXT impacting at 1232 ft/s.

The discretization error and other sources of experimental noise will find their way back into the analysis if the measured position or velocity are used in any of the downstream analysis. Consequently, all the subsequent analysis uses the model velocity and the time. The model position, x(t), is generated in the spreadsheet by numerically integrating the model velocity. The initial position (at impact) is zero. The model position for subsequent frames is the position of the previous frame plus the model velocity corresponding to the time of that frame times the time interval, dt (0.00005 s in this example). In other words, $x_{i+1} = x_i + V(t)\,dt$. The model acceleration is the change in the model velocity divided by the change in time between frames. The kinetic energy (in ft lbs) is $E = \tfrac{1}{2} mV^2$, where m is the mass of the bullet (converted to slugs, the proper English unit of mass).

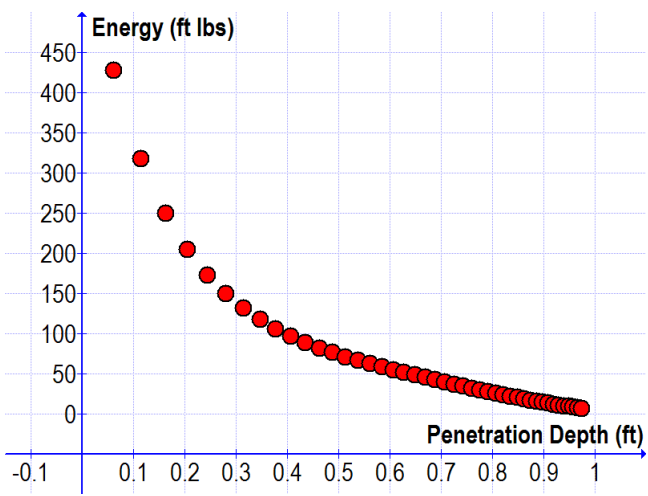

*Figure 2: Remaining kinetic energy vs. penetration depth for the 127 grain Winchester Ranger SXT bullet in 10% ballistic gelatin. Inspecting the graph suggests that the energy deposit in the first 15 cm would be close to 425 ft lbs – 75 ft lbs = 350 ft lbs. The energy deposit in the first 30 cm would be close to 425 ft lbs – 0 ft lbs = 425 ft lbs. More precise analysis is possible from interpolating the spreadsheet values for better estimates of the energy at any desired depth.*

There are then two ways to determine the retarding force, and having columns in the spreadsheet for both provides a double check on the analysis and also a sense for the magnitude of uncertainty being introduced with the numerical procedures. The first method to get the retarding force is Newton's second law, F = ma, where m is the bullet mass and a is the acceleration. Note that F = ma is a valid expression of Newton's second law only if the mass is constant. If the mass is changing, then there is a dm/dt term that needs to be estimated. High speed video kinematics is much simpler for bullets which do not fragment. If a is in ft/s/s and m is in slugs, the retarding force, F, is in pounds. The second method to get the retarding force is based on the Work-Energy theorem, F = dE/dx, where dE is the change in kinetic energy (the model energy, using the model velocity) between the current frame and the previous frame, and dx is the change in position between the current frame and the last, dx = Vdt. Conversions to metric units might then require additional columns in the spreadsheet.[2]

---

[2] When doing ballistics, American scientists are often torn between English units, where there tends to be a well developed intuition and feel both among American researchers and their likely audience, and Metric units (or MKS or SI or whatever). It is good to remember that the tyrants of the French revolution who imposed Metric units also beheaded Antoine Lavoiser, who was perfectly content measuring in pounds, ounces, gros, and grains. We recommend doing the main analysis in the

The resulting physical quantities in each column (energy, force, acceleration, etc.) can be plotted against either time or against distance (penetration depth). It is somewhat of academic interest to consider the time scales of the kinematic events, but most people gain more insight by plotting physical quantities against penetration depth. Figure 2 shows the remaining bullet energy as a function of penetration depth. The figure can be used quite simply to estimate values such as $E_{15}$ and $E_{30}$, which have been used to estimate incapacitation probabilities (Neades and Prather, 1991).

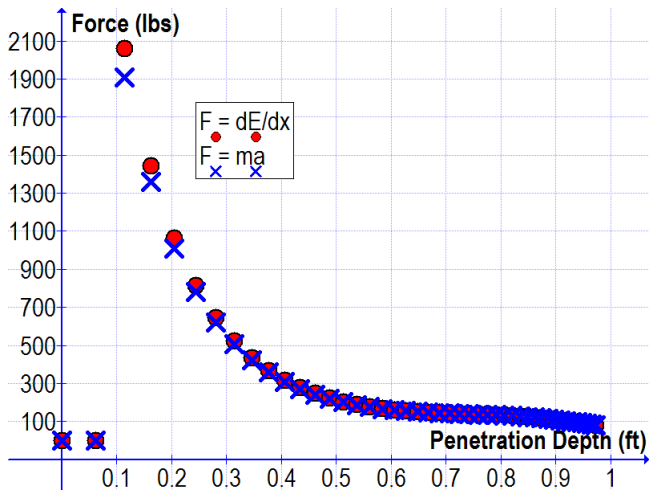

Figure 3: Retarding force vs. penetration depth for the 127 grain Winchester Ranger SXT in 9mm NATO determined from analysis of high speed video of a bullet penetrating 10% ballistic gelatin.

Figure 3 compares the retarding force curves obtained from the Work-Energy theorem (F = dE/dx) and Newton's second law (F = ma) by high speed video analysis. The two curves are within 10% of each other, which probably represents the level of error likely in the numerical procedure. Also note that the numerical method suggests that the retarding force is zero until the third frame. This is an artifact of the analysis procedure, and a better estimate would result from shifting the retarding force to the left one frame (which is done in the results section).

---

most comfortable set of units to subject each step to common sense assessments and then converting to SI units if needed to satisfy the unit tyrants or the intended audience. Good science does not depend on a particular system of units.

# Bullet Retarding Forces in Ballistic Gelatin by Analysis of High Speed Video

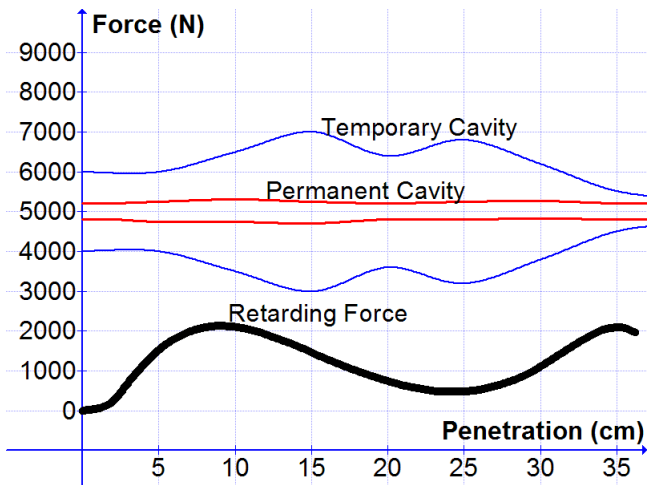

*Figure 4: Retarding force vs. penetration depth for the 124 grain FMJ in 9mm NATO. Note use of Metric units. The permanent and temporary cavities, determined from high speed video, are also shown.*

**Results**

The force curve as well as the temporary cavity and permanent cavity formed by the 124 grain full metal jacket (FMJ) 9mm NATO bullet are shown in Figure 4. This bullet is employed by many NATO nations for military service. It does not fragment or expand in soft tissue. Note the peak retarding force is only 2100 N which is 470 lbs. The relatively small peak retarding force is responsible for the modest permanent cavity volume and modest temporary cavity volume and also the modest ballistic pressure wave magnitude. Due to projectile tumbling, the retarding force has two peaks, corresponding to depths when the bullet is moving sideways. The temporary cavity also has two depths where its diameter displays local maxima. The energy deposit in the first 15 cm of penetration ($E_{15}$) is 172 ft lbs.

Peters (1990) describes a phenomena known as disk energy trading which explains why the peak temporary cavity diameters do not occur at the same depths as the peaks in the retarding force curves. The name "disk energy trading" refers to the technique of integrating volumes by the method of disks in Calculus. This mathematical method represents the temporary cavity as a series of disks of differing diameters, each with the diameter closely approximating the temporary cavity at a given penetration depth. In Peters' mathematical model "disk energy trading" describes the physical phenomena that the energy and retarding forces the bullet applies at one penetration depth is often transferred via stress and shear waves to other penetration depths.

# Bullet Retarding Forces in Ballistic Gelatin by Analysis of High Speed Video

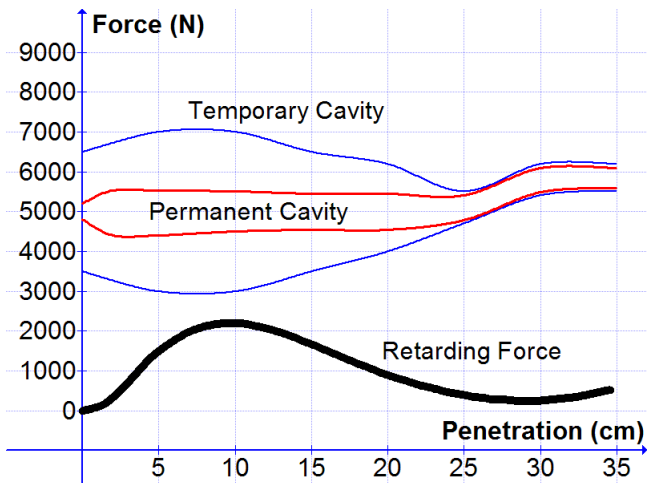

*Figure 5: Retarding force vs. penetration depth for the 147 grain WWB JHP load in 9mm NATO. The permanent and temporary cavities, also determined from the high speed video are also shown.*

Figure 5 shows the retarding force curve as well as the temporary cavity and permanent cavity determined by high speed video for the 147 grain WWB JHP load. This subsonic load was recommended for a time by the U.S. Federal Bureau of Investigation after the perceived penetration failure of a different bullet in a famous 1986 Miami shootout. Perceived weaknesses with this and other 9mm loads eventually led to the development and adoption of the .40 S&W. Note the peak retarding force is only 2200 N (close to 480 lbs), just slightly greater than the 124 grain FMJ. The relatively small peak retarding force is responsible for the modest permanent cavity volume and modest temporary cavity volume and also the modest ballistic pressure wave magnitude. The energy deposit in the first 15 cm of penetration ($E_{15}$) is 197 ft lbs.

Figure 6 shows the retarding force curve as well as the temporary cavity and permanent cavity determined by high speed video for the 147 grain Winchester Ranger SXT. Note the peak retarding force is 3600 N (over 800 lbs), much greater than the 124 grain FMJ or 147 grain WWB JHP. The larger retarding force is responsible for the significant increases in permanent cavity volume and the much larger temporary cavity. The energy deposit in the first 15 cm of penetration ($E_{15}$) is 240 ft lbs. This bullet is popular among law enforcement agencies because it is one of the more reliably performing 147 grain bullets in 9mm NATO.

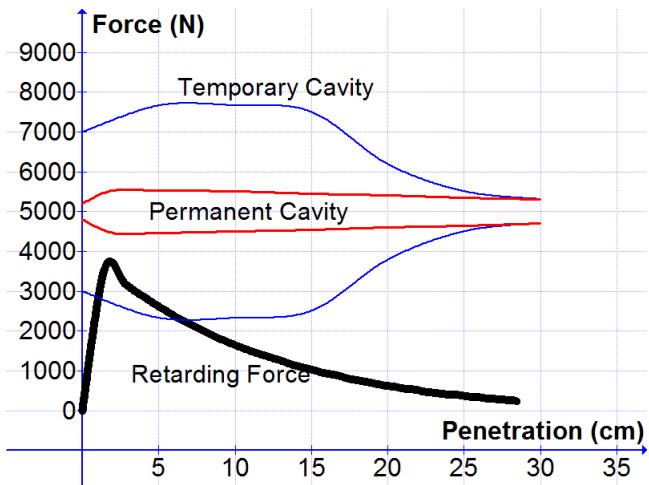

*Figure 6: Retarding force vs. penetration depth for the 147 grain Winchester Ranger SXT load in 9mm NATO. The permanent and temporary cavities are also shown.*

Retarding force, temporary cavity, and permanent cavity for the 127 grain Winchester Ranger SXT are shown in Figure 7. The peak retarding force is over 9000 N (over 2000 lbs), much greater than the forces of the other three 9mm NATO bullets considered here. The larger retarding force is responsible for the significant increases in permanent cavity volume and the much larger temporary cavity. The energy deposit in the first 15 cm of penetration ($E_{15}$) is 350 ft lbs. This bullet is popular among law enforcement agencies because of its outstanding terminal performance (for a pistol round).

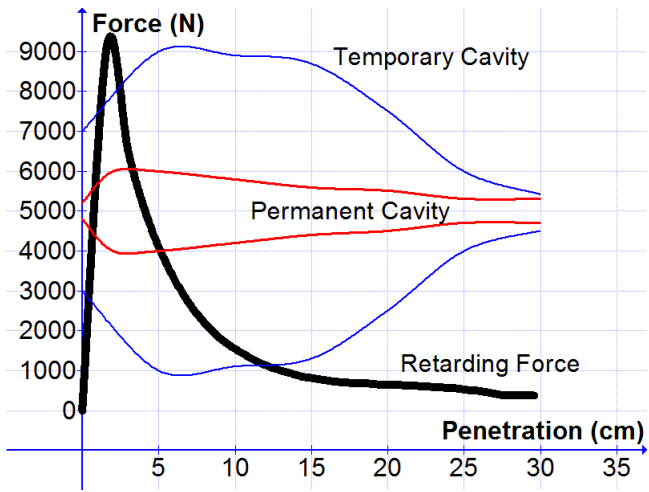

*Figure 7: Retarding force vs. penetration depth for the 127 grain Winchester Ranger SXT load in 9mm NATO. The permanent and temporary cavities are also shown.*

# Bullet Retarding Forces in Ballistic Gelatin by Analysis of High Speed Video

**Discussion**

Once the retarding force curve is computed, the stress fields and resulting shear can be computed with the mathematical model of Peters (1990). If the material properties of tissue are known, then the temporary stretch cavity and damage can also be predicted. The method of Lee et al. (1997) can be used to compute the ballistic pressure wave at any point in an isotropic medium, and with the appropriate element based model, the retarding force curve is the necessary input to computing the ballistic pressure wave throughout a more complex target in which the bullet path is well approximated by ballistic gelatin (soft tissue). Until now, the methods of Peters (1990) and Lee et al. (1997) could only be applied to non-deforming projectiles and the coefficient of drag as a function of velocity is generally not known very well. Determination of the retarding force curve with high speed video allows these methods to be applied to a much broader array of projectiles.

The method described above also allows determining the energy deposit in any desired range of penetration depths. This can be used to estimate the volume/diameter of wounded tissue as a function of penetration depth (Janzon, 1983). It can also be used to estimate the probability of incapacitation given a hit, P(I/H) on an enemy soldier (Neades and Prather, 1991). The details of the mathematical model BRL (later ARL) has used to estimate conditional incapation probabilities (also sometimes called $P_{kh}$) have changed slightly over the years, but are consistently based on knowing the energy deposit as a function of penetration depth. The approach in this discussion employs the energy lost by the bullet in the first 15 cm of penetration, $E_{15}$, and the model of Bruchey and Sturdivan (1968):

$$P(I/H) = 1 - e^{-a(E_{15})^n},$$

where the parameters a and n are determined by fitting to the incapacitation data on p. D-23 of "The Report of the M16 Rifle Review Panel: History of the M16 Weapon System" (M16, 1968).

| 9mm NATO Bullet | $E_{15}$ (ft lbs) | P(I/H) | M193 Equivalent Range (yards) |
|---|---|---|---|
| 124 FMJ | 172 | 0.330 | 620 |
| 147 WWB | 197 | 0.361 | 570 |
| 147 SXT | 240 | 0.410 | 490 |
| 127 SXT | 350 | 0.515 | 380 |

*Table 1: Conditional incapacitation probabilities for 9mm NATO bullets using the BRL method (Bruchey and Sturdivan, 1968; Neades and Prather, 1991). Also shown is the range where the M16 (using the M193 bullet) has an equivalent effectiveness (M16, 1968).*

Table 1 shows the resulting conditional incapacitation probabilities for 9mm NATO pistol bullets at very close range (< 10 yards). The 124 grain FMJ bullet has a P(I/H) of only 33%, which approximates the effectiveness of the M193 bullet fired from an M16 at 620 yards. Reports from Afghanistan show that the M16 family of rifles is rarely effective much beyond 300 yards (Ehrhart, 2009). It is therefore necessary and desirable to to employ small arms rounds with greater effectiveness than the currently fielded 124 grain FMJ in 9mm NATO, especially considering that pistols are most often employed in close quarters when the lingering threat of an unneutralized enemy combatant is much greater than when the enemy combatant is over 300 yards away. This unacceptable performance is consistent with the lackluster performance of 9mm NATO FMJ bullets in the Marshall and Sanow (2001) study as well as the long incapacitation times of 9mm NATO FMJ bullets on live goats in a laboratory study (Courtney and Courtney, 2007c).

# Bullet Retarding Forces in Ballistic Gelatin by Analysis of High Speed Video

The 147 grain WWB round in 9mm NATO has a conditional incapacitation probability of 36%, corresponding to an M16 bullet at 570 yards. This relatively poor performance is consistent with the poor performance of this round in the Marshall and Sanow (2001) one shot stop study, an epidemiological type study of bullet effectiveness stopping agressive human attackers. The prediction of poor effectiveness is also consistent with the relatively long incapacitation time in a laboratory experiment in human sized goats (Courtney and Courtney, 2007c), as well as the long distances run by deer after being shot through the lungs with this bullet in a carefully controlled field study (Courtney and Courtney, 2007d).

The 147 grain Winchester Ranger SXT round in 9mm NATO has a conditional incapacitation probability of 41%, corresponding to an M16 bullet at 490 yards. The rank ordering of these three bullets (124 grain FMJ, 147 grain WWB, and 147 grain SXT) by P(I/H) in Table 1 is the same as the rank ordering of the full metal jacket, 147 grain WWB and 147 grain SXT in the Marshall and Sanow (2001) one shot stop study and also the same as the rank ordering of average incapacitation times in the laboratory experiment with human sized goats (Courtney and Courtney, 2007c).

The 127 grain Winchester Ranger SXT round in 9mm NATO has a conditional incapacitation probability of 52%, corresponding to an M16 bullet at 380 yards. This level of performance is not particularly impressive compared with what is possible from rifles at close range. For example, Dean and LaFontaine (2008) report that all rifle rounds tested have conditional incapacitation probabilities above 80% inside of 20 yards and conditional incapacitation probabilities above 90% are possible.[3] This is consistent with the findings of Marshall and Sanow (2001) that the rifle loads in their epidemiological type study have one shot stop ratings above 90%. However, the conditional incapacitaion probability of 52% for the 127 grain SXT is probably near the upper end of what is possible given the energy limitations of the 9mm NATO pistol cartridge, especially if confined to considering bullets which do not fragment and meet the FBI minimum penetration requirements of 12" in calibrated ballistic gelatin. With an expected effectiveness approximating an infantry rifle bullet at 380 yards, this level of performance from a pistol bullet is necessary given the military requirements for close quarters battle in which pistols are most commonly employed.

In summary, this paper has presented the details of a method employing high speed video to quantify the kinematics of bullets penetrating ballistic gelatin, including determination of retarding forces and energy deposit. This method has been demonstrated with an example studying the terminal performance of 4 different bullet designs in 9mm NATO resulting in estimates of bullet effectiveness. These estimates of bullet effectiveness yielded by the method are consistent with other studies.

**Acknowledgements**
This research was supported in part by BTG Research (www.btgresearch.org ) and by the United States Air Force Academy. Distribution A. Approved for public release. Distribution unlimited. The views expressed in this paper are those of the authors and do not necessarily represent those of the U.S. Air Force Academy, the U.S. Air Force, the Department of Defense, or the U.S. Government.

---

[3]This conclusion requires triangulating a bit with respect to the interpretation of Figure 8 in Dean and LaFontaine (2008), since the vertical axis in Figure 8 is unlabeled. The authors of the present study believe that ARL continues to use one of the conditional probability incapacitation models described in Neades and Prather (1991) or a subsequent refinement and that the each horizontal grid line in Figure 8 represents an increase of 10% in effectiveness. The oscillations in the better performing single shots are due to enhanced energy transfer early in the penetration due to the yaw angle at impact. Damping of the oscillations over the first 50 m is due to the bullets "going to sleep" (damping of pitch and yaw). Decreasing effectiveness over the first 50 m is due to a combined decreasing impact energy along with lower energy deposit early in the penetration. We have not been able to reconstruct whether the effectiveness of controlled pairs was estimated simply by summing the energy deposit of two shots and then using an energy deposit formula of Neades and Prather (1991) or subsequent refinement or whether the effectiveness of controlled pairs has been estimated by combining the conditional incapacitation probabilities of two single shots by the rules of probability. We suspect the former, but believe using the rules of probability would be more appropriate.

**References**

Bellamy RF, Zajtchuk R. The physics and biophysics of wound ballistics. [ed.] Zajtchuk R. *Textbook of Military Medicine, Part I: Warfare, Weaponry, and the Casualty, Vol. 5.* Washington, D.C.: Office of the Surgeon General, Department of the Army, 1990.

Beyer JC, ed. *Wound Ballistics in World War II.* Medical Department, United States Army. Washington, D.C. : Office of the Surgeon General, Department of the Army, 1962. Available at: http://www.dtic.mil/cgi-bin/GetTRDoc?AD=ADA291697&Location=U2&doc=GetTRDoc.pdf . Accessed July 15, 2012.

Bruchey WJ Jr., Sturdivan LM. An Instrumented Range Meeting the Requirements of a Wound Ballistics Small Arms Program. BRL-TN-1703,U.S. Army Ballistic Research Laboratory, Aberdeen Proving Ground, MD, 1968.

Courtney A, Courtney M. Physical Mechanisms of Soft Tissue Injury from Penetrating Ballistic Impact. 2012. United States Department of Defense.

Courtney AC, Courtney MW. Links between traumatic brain injury and ballistic pressure waves originating in the thoracic cavity and extremities. *Brain Inj.* 2007a; 21:657-662.

Courtney M, Courtney A. Relative incapacitation contributions of pressure wave and wound channel in the Marshall and Sanow data set. Cornell University Library, Medical Physics. 2007b.

Courtney M, Courtney A. Ballistic pressure wave contributions to rapid incapacitation in the Strasbourg goat tests. Cornell University Library, Medical Physics. 2007c.

Courtney M, Courtney A. A Method for Testing Handgun Bullets in Deer. Cornell University Library, Medical Physics. 2007d.

Dean G, LaFontaine D. Small Caliber Lethality: 5.56 mm Performance in Close Quarters Battle. WSTIAC Quarterly, Vol. 8, No. 1. 2008.

Ehrhart TP. Increasing Small Arms Lethality in Afghanistan Taking Back the Infantry Half-Kilometer. ARMY COMMAND AND GENERAL STAFF COLL FORT LEAVENWORTH KS SCHOOL OF ADVANCED MILITARY STUDIES, 2009.

Janzon, B. PhD Thesis on "High Energy Missile Trauma - a Study of the Mechanisms of Wounding of Muscle Tissue" 1983.

Krajsa J. Příčiny vzniku perikapilárních hemoragií v mozku při střelných poraněních (Causes of pericapillar brain haemorrhages accompanying gunshot wounds). Brno: Institute of Forensic Medicine, Faculty of Medicine, Masaryk University, 2009. http://is.muni.cz/th/132384/lf_d/Autoreferat.pdf Accessed September 20, 2012.

Krauss M, Miller JF. Studies in Wound Ballistics: Temporary Cavities and Permanent Tracts Produced by High-Velocity Projectiles in Gelatin. Chemical Warfare Laboratories, U.S. Army. Washington, D.C.: Defense Technology Information Center, 1960; AD233840. Available at: http://www.dtic.mil/cgi-bin/GetTRDoc?AD=AD233840&Location=U2&doc=GetTRDoc.pdf . Accessed July 12, 2012.

Lee M, Longoria RG, Wilson DE. Ballistic waves in high-speed water entry. *J Fluids Struct.* 1997; 11:819-844.

Marshall E, Sanow E. Stopping Power. Paladin Press, 2001.

The M16 Rifle Review Panel. History of the M16 Weapon System. 1968. United States Department of Defense. ADA953110.

Neades DN, Prather RN. The modeling and application of small arms ballistics. United States Department of Defense, ADA240295, 1991.

Peters, Carroll E. A mathematical-physical model of wound ballistics. *J Trauma (China)* 1990; 2(6, Supp):303-318.

Selman YS, et al., Medico-legal Study of Shockwave Damage by High Velocity Missiles in Firearm Injuries, Fac Med Baghdad 2011; Vol. 53, No. 4.

Suneson A, Hansson HA, Seeman T. Pressure wave injuries to the nervous system caused by high-energy missile extremity impact: part I. local and distant effects on the peripheral nervous system. A light and electron microscopic study on pigs. *J Trauma.* 1990a; 30:281-294.

Suneson A, Hansson HA, and Seeman T. Pressure wave injuries to the nervous system caused by high energy missile extremity impact: part II. distant effects on the central nervous system. A light and electron microscopic study on pigs. *J Trauma.* 1990b; 30:295-306.